# Strengthened proximity effect at grain boundaries to enhance inter-grain supercurrent in $Ba_{1-x}K_xFe_2As_2$ superconductors

Zhe Cheng[1], Chiheng Dong[1,4,*], Huan Yang[2], Qinghua Zhang[3], Satoshi Awaji[5], Lin Gu[3], Hai-Hu Wen[2,*], Yanwei Ma[1,4,*]

[1]*Key Laboratory of Applied Superconductivity, Institute of Electrical Engineering, Chinese Academy of Sciences, Beijing 100190, China*
[2]*Center for Superconducting Physics and Materials, National Laboratory of Solid State Microstructures and Department of Physics, Nanjing University, Nanjing 210093, China*
[3]*Beijing National Laboratory for Condensed Matter Physics, Institute of Physics, Chinese Academy of Sciences, Beijing 100190, China*
[4]*University of Chinese Academy of Science, Beijing, 100049, China*
[5]*High Field Laboratory for Superconducting Materials, Institute for Materials Research, Tohoku University, Sendai 980-8577, Japan*
\* E-mail: *dongch@mail.iee.ac.cn, hhwen@nju.edu.cn, ywma@mail.iee.ac.cn*

**ABSTRACT**

Iron-based superconductors have great potential for high-power applications due to their prominent high-field properties. One of the central issues in enhancing the critical current density of iron-based superconducting wires is to reveal the roles and limitations of grain boundaries in supercurrent transport. Here, we finely tuned the electronic properties of grain boundaries by doping $Ba_{1-x}K_xFe_2As_2$ superconductors in a wide range ($0.25 \leqslant x \leqslant 0.598$). It is found that the intra-grain $J_c^{intra}$ peaks near $x \approx 0.287$, while the inter-grain $J_c^{inter}$ has a maximum at about $x \approx 0.458$. Remarkably, the grain boundary transparency parameter defined as $\varepsilon = J_c^{inter}/J_c^{intra}$ rises monotonically with doping. Through detailed microscopic analysis, we suggest that the FeAs segregation phase commonly existing at grain boundaries and the adjacent grains constitute superconductor-normal metal-superconductor (SNS) Josephson junctions which play a key role in transporting supercurrent. A sandwich model based on the proximity effect and the SNS junction is proposed to well interpret our data. It is found that overdoping in superconducting grains largely strengthens the proximity effect and consequently enhances the intergrain supercurrent. Our results will shed new insights and inspirations for improving the application parameters of iron-based superconductors by grain boundary engineering.

## 1. Introduction

Today's high-field superconducting equipments require kilometer-long superconducting wires with a prominent ability to carry loss-less current. The physical properties of the wires determined by the inherent crystal and electronic structures of the inner superconducting materials lay a fundamental restriction on their practical applications. Iron-based superconductors [1] (IBSC) discovered in 2008 perfectly meet the basic application requirements and may provide a new opportunity to fill the gap in current superconducting technology [2–5]. After decades of research and development, iron-based superconducting long tapes fabricated by the powder-in-tube (PIT) method have already been made into superconducting coils, indicating a great potential to be applied in high field magnets [6–8]. However, the highest critical current density $J_c$ ever reached in $Ba_{1-x}K_xFe_2As_2$ PIT tapes is still much smaller than the $Ba_{1-x}K_xFe_2As_2$ films. [9]. The key factors controlling the inter-grain transport of supercurrent continue to be controversial and subjects of intense interests. Nowadays, great endeavors are mainly contributed to increasing the compactness of grains [10–12] and suppressing the out-of-plane misorientations [13]. Remarkable improvement of the $J_c^{transport}$ surpassing the level for practical applications has been made by varieties of methods [14]. However, the limitation of these techniques becomes more evident in recent years when we try to enhance the $J_c$ to a new level. It is intriguing to reconsider the key obstacles to the improvement of $J_c^{transport}$ from different prospects.

As a model system for IBSC conductors, the $Ba_{0.6}K_{0.4}Fe_2As_2$ is widely accepted as the promising one in making superconducting tapes with the highest $T_c$ to achieve the best performance [15,16]. However, recent studies show that the highest $J_c$ of the $Ba_{1-x}K_xFe_2As_2$ (BaK122) single crystals appears at the slightly underdoped level [17]. The orthorhombic/antiferromagnetic (AFM) domain boundaries that may still exist in the underdoped samples are considered to be the main source of the enhanced flux pinning [18]. This result strongly implies that the BaK122 PIT tapes could achieve higher $J_c$ if we finely tune the chemical composition to the underdoped side. However, the supercurrent crossing the interface between adjacent crystallites is largely complicated by the microscopic electronic, structural and chemical variations at grain boundaries [19]. It is therefore intriguing to perform an extensive research on the doping dependence of $J_c$ of the BaK122 tapes to reveal whether a high intragrain supercurrent can be maintained in polycrystalline tapes and gain a detailed understanding of the doping induced evolution of grain boundary characteristics.

In this work, we fabricated stainless steel (SS)/silver sheathed $Ba_{1-x}K_xFe_2As_2$ composite tapes (Supplementary materials, Fig. S1) with K content in region $0.25 \leq x \leq 0.598$ by the PIT method and study the doping dependence of superconductivity. It is found that the intragrain critical current density, $J_c^{intra}$, follows the same doping dependence as that of the $Ba_{1-x}K_xFe_2As_2$ single crystals. Unexpectedly, the intergrain critical current density $J_c^{inter}$ reaches the maximum in the slightly overdoped region. Moreover, the grain boundary transparency parameter $\varepsilon = J_c^{inter}/J_c^{intra}$ increases monotonically with $x$, which can be described by the proximity-induced SNS Josephson junction model. We suggest that the FeAs segregation phase at grain boundaries is the main factor governing the transport critical current density of the $Ba_{1-x}K_xFe_2As_2$ superconducting tapes.

## 2. Materials and Methods

*2.1 Sample preparation.*

The precursor powders were synthesized by the solid-state-reaction method. High-purity Ba (99 wt.%), K (99.95 wt.%), Fe (99.95 wt.%) and As (99.9999 wt.%) were used as the raw materials. The previously prepared BaAs and KAs intermediates were thoroughly mixed with Fe and As powders via a ball milling machine according to the nominal compositions $Ba_{1-x}K_xFe_2As_2$ ($x$=0.25, 0.3, 0.4, 0.46, 0.48, 0.5, 0.56, 0.6). All the raw materials were handled in an Ar filled glove box (< 0.1 ppm $O_2$ and < 0.1 ppm $H_2O$) to reduce contamination. The mixture was sintered at 900 °C for 35 hours. The precursor bulks were ground into powders and filled into a Ag tube (outer-diameter 8mm, inner-diameter 5 mm). The Ag tube was then drawn and flat-rolled into tapes which can be inserted into the pre-rolled stainless tube. The composite was flat rolled into the tapes with thickness $t$~0.8 mm. The tapes were finally sintered at 750 °C for half an hour. This fabrication procedure was applied to all the tapes with different doping levels. The thoroughly mixed powders of Fe and As with a ratio of 1:1 were pressed into a pellet and sintered at 700 °C for 24 hours to obtain the FeAs polycrystalline bulks.

*2.2 Characterization methods.*

The crystal structure was characterized by X-ray diffraction (Bruker D8 Advance) measurements at room temperature. The actual chemical composition of the tape was examined on at least 10 points of the sample by the electron probe microanalysis (EPMA; JEOL JXA8230). We utilized the electron backscattered diffraction orientation imaging microscopy (EBSD-OIM) to examine the grain texture of the superconducting core. Microscopic characterization of the grain boundaries was performed on a spherical aberration corrected transmission electron microscope (JEOL JEM-ARM200F). The composition variation across the grain boundary was examined by the energy dispersive spectroscopy (EDS) equipped on the transmission electron microscope. The transport critical current $I_c$ was measured in liquid helium (4.2 K) up to 14 T by the standard four-probe method with the criterion of 1 μV/cm. The field is applied parallel to the tape surface during the $I_c$ measurements. Magnetic and electrical properties of the superconducting cores were measured on a Physical Property Measurement System (PPMS-9, Quantum Design) after peeling off the metal sheaths. The Hall effect is measured by the van der Pauw method. The superconducting cores were shaped into rectangular slabs before the resistivity and Hall effect measurements.

## 3. Results and Discussion

Fig. 1a shows the powder X-ray diffraction (XRD) patterns of the samples with different doping levels. All the diffraction peaks can be indexed with the *I4/mmm* space group. No sign of impurity phase is observed. To obtain detailed crystal structure information, we performed Rietveld refinements to the XRD patterns (Fig. S2). The lattice parameters from fitting are shown in Fig. 1b. Through K doping, the *c*-axis lattice constant elongates, while the parameters in the *ab*-plane decrease. The lattice parameters depend linearly on the K content. These results are similar to that of the published papers [20]. The actual K contents measured by electron probe microanalysis (EPMA) are close to the nominal ones, as shown in Table.S1 (Supplementary materials). We will use the measured K contents in the following part of the paper.

The metal sheaths were peeled off to obtain the superconducting cores as the studied samples. The temperature dependence of the normalized susceptibility of the superconducting core is depicted in Fig. 1c. All samples are measured under a zero-field-cooling (ZFC) procedure with H=5 Oe parallel to the tape plane. The $T_c^{mag}$ is determined from the point at which the magnetization starts to deviate from the normal-state linear background. The sharp superconducting transition near $T_c^{mag}$ indicates a homogeneous superconducting state. As we expect, the sample with *x*=0.397 possesses the highest $T_c$=38 K. Under- or over-doping suppresses $T_c^{mag}$ but causes no conspicuous broadened superconducting transition. Fig. 1d and e shows the temperature dependence of resistivity of the superconducting core. Clearly, all the samples show a metallic behavior above $T_c$. In addition, the resistivity at a fixed temperature in the normal state gradually shifts to the lower values with the increased K content. The doping dependence of the measured $T_c$ is summarized in the phase diagram shown in Fig. S3. The consistency between our data and the published ones (blue line) again validates the accuracy of the measured K content [21].

Fig.2 shows the field dependence of the critical current density at liquid helium temperature. We choose 4.2 K as the studied temperature instead of the reduced temperature $t=T/T_c$ because the variation of $T_c$ is small in the doping range, e.g. $T_c^{mag}$ ranges from 32.5 K to 37.6 K within 0.287<$x$<0.551. In order to study the influence of doping on the intra-grain critical current density $J_c^{intra}$, we thoroughly ground the superconducting cores into powders and measured the isothermal magnetization hysteresis loops at 4.2 K. The average size of the powder particles is evaluated by the SEM. According to the Bean critical state model, the $J_c^{intra}$ can be estimated by $J_c^{intra}=30\Delta M/d$, where $\Delta M$ is the magnetization difference (emu/cm$^3$) between the field ascending and descending branches of the loop, and $d$ is the average grain size in cm [22]. Fig. 2a shows the field dependence of $J_c^{intra}$. Similar to the results obtained in the BaK122 single crystals [17], the sample with $x$=0.287 achieves the highest $J_c^{intra}$. It indicates that the flux pinning force is stronger in the grains with $x$=0.287. On the over-doped side where $x$ > 0.5, the $J_c^{intra}$ is considerably depressed.

For the measurements of the transport critical current $I_c$, at least four samples were examined for each doping level to guarantee reproducibility. The good protection provided by the strong SS sheath and the homogeneous cold deformation procedures cooperatively lead to the uniform and close values of $I_c$ for the tapes with the same composition. We denote the transport $J_c$ as the average inter-grain critical current density, $J_c^{inter}$. Fig. 2b summarizes the field dependence of $J_c^{inter}$ at 4.2 K for the best tapes at different doping levels. One can see that $J_c^{inter}$ presents a disparate $x$ dependence from $J_c^{intra}$. The tape with $x$=0.287 which achieves the best $J_c^{intra}$ no longer exhibits the highest $J_c^{inter}$. The optimally doped tape ($x$=0.397) achieves a $J_c^{inter}$=7.3 × 10$^4$ A/cm$^2$ at 10 T, which is close to the reported value [23], indicating the good reliability of the fabrication technology. Surprisingly, the tape with $x$=0.458 achieves the best performance of $J_c^{inter}$. The $J_c^{inter}$ of this sample reaches $1.03 \times 10^5$ A/cm$^2$ at 10 T, which has exceeded the level for practical applications. Further increase of the doping level beyond $x=0.48$ results in a gradually decreased $J_c^{inter}$.

Fig. 3a compares the doping dependence of the inter- and intra-grain critical current density at 4.2 K and 8 T. The highest $J_c^{intra}$ locates at the doping level near $x=0.287$. On the contrary, the $J_c^{inter}$ climbs continuously with doping up to the peak near $x=0.458$. This discrepancy demonstrates that K doping not only changes the intra-grain flux pinning property but also varies the current transport efficiency across the grain boundaries. It is noticed that there is a sudden drop of $J_c^{intra}$ and $J_c^{inter}$ when $x$ increases to 0.496. Based on the similar trends of the curves, we suggest the decrease of $J_c^{inter}$ is caused by the suppressed $J_c^{intra}$. Ishida *et al.* found that the pinning mechanism of Ba$_{1-x}$K$_x$Fe$_2$As$_2$ single crystals changes from $\delta l$ to $\delta T_c$ pinning when $x>0.51$ [24]. We suggest that the abrupt decline of $J_c^{intra}$ is caused by the variation of pinning properties of the superconducting grains. We define the grain boundary transparency parameter $\varepsilon=J_c^{inter}/J_c^{intra}$ to describe the supercurrent transport efficiency across the grain boundaries (GBs). The black squares and red circles in Fig. 3b are the GB transparency parameters $\varepsilon$ at 6 T and 8 T, respectively. They nearly overlap with each other, implying that $\varepsilon(x)$ follows the same doping dependence at different magnetic fields. On the under-doped side with $x=0.287$, only 3.8 % of the intra-grain supercurrent is preserved across the GBs, despite the better intra-grain flux pinning. Further doping to $x=0.397$ marginally improves the $\varepsilon$ to 8.3 %. On the slightly overdoped side where $x=0.458$, the GB transparency parameter increases to $\varepsilon\sim12.5$ %. In the meantime, the $J_c^{inter}$ rises to the peak shown in Fig. 3a. Interestingly, further doping induces a continuous increase of $\varepsilon$. The GB transparency parameter even reaches 46.8 % at $x=0.598$.

It is important to note that the density and texture of the superconducting filaments should underlie the current carrying ability of iron-based superconducting wires and tapes. The careful analysis shown in Fig. S4 indicates that the density and texture of the filaments are dominated by the fabrication process which was maintained in the same way for all samples with different doping levels. The incongruence between the doping dependences of $\varepsilon$, density and texture inspires us to explore the intrinsic reasons for the improvement of the GBs transparency with K doping.

To scrutinize the characteristics of the GBs, we performed a detailed investigation on the microstructure of the grain boundaries in the under-, optimally- and over-doped samples by a transmission electron microscope (TEM). Fig. 4a-c show the bright-field TEM images of the grain boundary network observed from the direction perpendicular to the tape plane. The plate-like grains with polygon shapes are separated by the GBs with different thicknesses $t$. As marked in (a-c), we define the type I and II GBs as the ones with $t>5$ nm and $t\leqslant 5$ nm, respectively. Fig. 4d-e are the enlarged views of the typical type II GBs. While Fig. 4f presents the type I GB that separates the superconducting grains more widely. The obvious difference in contrast demonstrates a secondary phase agglomerating at GBs. It must be emphasized that both types of GBs can be found randomly in the tapes with different doping levels. For a quantitative evaluation, we perform statistical analysis on the grain size and the grain boundary thickness. The grain size in (g) is about 300-600 nm, regardless of the doping level. As shown in (h), the GBs thickness $t$ ranges from 2 nm to 50 nm. The $t$ distribution of the three doping levels nearly overlaps with each other, suggesting no obvious doping dependence of the distribution of the GBs thickness. It is worth noting that the fraction of the type II GBs is higher than 40 %. Fig. 4i shows an energy dispersive spectroscopy (EDS) line scan across a GB in the inset of (h) (red line). There is no evident variation of Ba and K levels when crossing the GB. On the contrary, a sudden increase of Fe and As appears at the grain-GBs interface. It strongly suggests that the GBs are mainly contaminated by the FeAs wetting phase.

FeAs is widely noted as a concomitant phase during the high temperature sintering of iron-arsenide superconductors [25], whose formation mechanism is still beyond present understanding. It agglomerates at GBs and is sandwiched by the superconducting phases. It turns the system into a SNS Josephson junction network. To better understand the properties of the SNS junction, we separately studied the electrical characteristics of the FeAs and BaK122 phases. As shown in Fig. 5a, the under-doped samples show a fast increase of $R_H$ below 200 K. This escalation is gradually suppressed with further doping. There appears a broad peak near 100 K for the samples with $x>0.458$, which is also observed in the $Ba_{1-x}K_xFe_2As_2$ single crystals [26]. We calculated the effective hole density at 50 K by $n_{eff}=1/R_He$, where $e$ is the elementary charge, and compare it with the data from the references [26,27], as shown in (b). The consistency of our data to the others confirms the reliability. The $n_{eff}$ continues increasing from $2.21\times 10^{21}$ cm$^{-3}$ at $x=0.25$ to nearly $10^{22}$ cm$^{-3}$ at $x=0.598$, which is close to the level of low-temperature superconductors [28] (LTSC). As proved by the angle-resolved photoemission spectroscopy experiments, the electronic band structure of Ba$_{1-x}$K$_x$Fe$_2$As$_2$ progressively varies with K doping, leading to the largely expanded hole pockets and the contracted electron pockets [29,30]. Moreover, the residual resistance ratio (RRR) in Fig. 5c and the conductivity σ at 50 K shown in Fig. 3 corroborate that the metallicity of the BaK122 becomes better at a higher doping level. The simultaneous enhancement of $n_{\text{eff}}$, σ and ε in Fig. 3b implies that the GBs transparency is related to the electronic state of the BaK122 grains if we assume that the weak-link region is constructed by the FeAs phase. The temperature dependence of the resistivity and the Hall coefficient in Fig. 5d demonstrates that the FeAs behaves as a normal metal with electrons as the dominant charge carriers.

When the FeAs segregation phase is in close contact with the superconducting grains, as shown by the images in Fig. 4, we may model these GBs as the SNS junctions involved with the proximity effect. Therefore, the superconducting order parameter $|\psi|$ extends to the normal metal (NM), here the FeAs region, with a characteristic length $\xi_n$, as sketched in Fig. 6a. The enhanced metallicity of the superconducting grains strengthens the proximity effect and finally induces the overlap of the order parameter from both sides [31,32]. To quantitatively describe this phenomenon, the normalized order parameter $|\psi_{NM}|/|\psi_S|$ inside a 6 nm thick FeAs layer is simulated based on the theory reported in refs. [32,33], as shown in the inset of Fig. 6b. It is obvious that the order parameter in the FeAs layer is gradually enhanced by K doping. We further consider the GB transparency by taking the critical current density $J_c^{\text{inter}}$ crossing a GB with thickness $t$ over the intra-grain critical current density $J_c^{\text{intra}}$ in a one-dimensional SNS Josephson junction [32]:

$$\varepsilon(t) = \frac{J_c^{\text{inter}}}{J_c^{\text{intra}}} \approx \frac{J_{Dn}}{J_{Dsc}} \approx \frac{3\sqrt{3}\rho_s\xi_s}{\rho_n\xi_n}\left\{\sqrt{\left(\frac{\rho_s\xi_s}{\sqrt{2}\rho_n\xi_n}\right)^2 + 1} - \frac{\rho_s\xi_s}{\sqrt{2}\rho_n\xi_n}\right\}^2 \exp\left(-\frac{t}{\xi_n}\right), \quad (1)$$

where $J_{Dn}$ is the depairing critical current density of the NM layer, $J_{Dsc} = \frac{\phi_0}{3\sqrt{3}\pi\mu_0\xi_s\lambda_s^2}$ is the depairing critical current density of the S layer [34], $\phi_0$ is the flux quantum, $\mu_0$ is the vacuum permeability, $\xi_s$ is the coherence length of BaK122 derived from the magnetoresistance measurements (Fig. S5), $\xi_n$ is the decay length of $|\psi|$ in the FeAs region. Here $\rho_s$ and $\rho_n$ are the normal state resistivities of the BaK122 and the FeAs, respectively. The $\xi_s$ and $\rho_s$ depend on the doping state of the grains. For a fixed thickness $t$ of the FeAs sandwiched layer, $\rho_n$, $\xi_n$ can be treated as constants. We propose a current path model of the SNS GB network, as shown in Fig. S6. Considering the distribution of the GB thickness shown in Fig. 4h, we sum $\varepsilon(t)$ over the GBs with different thicknesses, $\varepsilon = \sum_i \varepsilon_i(t_i)f_i(t_i)$, where $f_i(t_i)$ is the fraction of the GB with thickness $t_i$. We input the determined values of the parameters $\rho_s$, $\xi_s$ into the above-mentioned equation and only set $\xi_n$ free to fit the experimental data, as shown in Fig. 6b. The comparison between the theoretical simulations and the experimental results shows quite good consistency, which justifies our proposed model. The fitted $\xi_n$ is about 2.72 nm. This value is valid for all the doping levels. We roughly estimate the coherence length of FeAs in dirty limit ($\xi_n \gg l$, $l$ is the mean free path). The $\xi_n$ can be described by $\xi_n \approx \sqrt{\frac{\hbar D_n}{2\pi k_B T}}$, where $k_B$ is the Boltzmann constant, $\hbar$ is the reduced Planck constant, $D_n = v_F l/3$ is the diffusion coefficient, $v_F$ is the Fermi velocity of the NM layer (here FeAs) [35]. According to this equation, the $\xi_n$ depends on the temperature and the properties of the FeAs layer. The mean free path of FeAs is roughly estimated by $l = v_F \tau$, where $\tau = \frac{m_0}{\rho n e^2}$ is the collision time, $m_0$ and $e$ are the mass and charge of an electron, $\rho$ and $n$ are the resistivity and the charge carrier density at low temperatures. On the hypothesis that the $v_F$ of FeAs [36] is about $10^5$ m/s, the $\xi_n$ is estimated to be larger than 2.4 nm below 10 K, which is close to the fitting results. Since the data of both the inter- and intra-grain critical current densities were determined under a high magnetic field, thus in principle we need to calculate the transparency $\varepsilon$ using a formula with magnetic fields. This formula was given by Carty *et al.* [32]. Detailed analysis finds that, the formula with a field is different from eq.(1), but with a fixed magnetic field and related values involved, the in-field equation will reduce to eq.(1) in terms of the fitting parameters of $\rho_s$, $\xi_s$. The model, although crude in itself, can fit the data quite well. This indicates that most of the GBs in the tapes are constructed by the FeAs segregations and they form the SNS type Josephson junctions network.

The electron-phonon mediated superconductors such as the LTSC and MgB$_2$, where the coherence length is usually much larger than the intergrain distance, have strongly coupled GBs. However, IBSC share many commonalities with the copper-oxide superconductors. From a fundamental point of view, the IBSC possess high upper critical fields and normal-state resistivities, small coherence lengths $\xi$, low charge carrier densities $n$ and Fermi energy $E_F$, *etc.*, these cooperatively result in more easily depressed superconducting order parameter at GBs [19,37]. In the case of strongly coupled GBs with no secondary phase, the inter-grain $J_c$ may be suppressed by interface charging and band bending effect [38]. This situation can be largely ameliorated by overdoping calcium or oxygen [39–41]. Song *et al.* found strong Ca segregation at GBs which reduces the GB charge and the screening length and consequently increases the inter grain supercurrent [42]. Indeed, better performance of the cuprates practical conductors is achieved in the overdoped Bi2212 wires and the YBCO films [43,44]. Similarly, the $J_c^{inter}$ of the cuprates practical conductors increases monotonically with the hole density [39,43]. However, the situation for the IBSC is different. The GBs in IBSC tapes are sandwiched with the FeAs wetting phase. It is therefore reasonable to observe a low GB transparency parameter. The astonishing enhancement of $\varepsilon$ over 40 % at the overdoped state cannot be ascribed to the GB doping by K atoms, which is not observed by the EDS analysis, but to the strengthened proximity effect in the SNS Josephson junction.

In the past decade, the labyrinth of boosting the IBSC conductor performance revolves around the texture and density of the filaments. Our studies highlight the necessity for improving the GBs properties beyond the aforementioned factors. First of all, the secondary phases between grains largely expand GBs thickness and adversely affect the intergrain supercurrent. The FeAs wetting phase which forms near 700 ºC is the archenemy to grain boundary transportation. Undoubtedly, the most effective way to improve the transport critical current density is just to reduce the amount of the wetting FeAs phase, in order to diminish the junction area. Successful experience of enhancing the $J_c$ of Bi2212 wires by removing the amorphous blocking layer between adjacent grains also encourages us to, at least, partly restore the superconducting channels by diminishing the contaminated GBs of IBSC tapes [45]. However, the formation mechanism of the FeAs phase during the solid-state-reaction has not yet been fully understood, leaving GBs far from optimization. If this problem can be overcome by some modified fabrication processes, the advantage of the strong intra-grain flux pinning in the slightly under-doped tapes will be well taken to achieve a high $J_c$. In the meantime, it is found that the sandwich structure of GBs can be considered as proximity-induced SNS Josephson junctions in the polycrystalline tapes. They may be modified by adjusting the charge carrier density in the grains, thus changing the GBs into strong pinning centers. Although the transparency parameter reaches the highest value in the overdoped samples, the $J_c^{inter}$ is appreciably limited by the $J_c^{intra}$ which is determined by the superconductor itself. The best performance achieved in the slightly over-doped sample results from the counterbalance between the GB transparency and the intragrain current. Furthermore, the sample with $x$=0.458 shows an almost undisturbed $T_c$, a high upper critical field and large $n$-values determined by I $\propto$ V$^n$ (where I is the applied current, V is the measured voltage, as shown in Fig. S1b), indicative of good potential to be applied at high fields. We suggest that this composition can be employed in the future fabrication of iron-based superconducting long wires. Nevertheless, there is still considerable space for improving the current-carrying ability of IBSC conductors. Our extensive studies presented here and the clear demonstration of the SNS junction like grain boundaries certainly shed new light on understanding the limiting factors for the transport critical current densities in IBSC tapes. This will stimulate a comprehensive understanding of the microstructural and electronic properties of the GBs, and is important to pave a road for exploring the large-scale applications of iron-based superconducting wires and tapes.

## 4. Conclusion

Doping dependence of the critical current density of Ba$_{1-x}$K$_x$Fe$_2$As$_2$ superconducting tapes with a wide doping range (0.25$\leq x \leq$0.598) is systematically studied. Surprisingly, the grain boundary transparency parameter described by $\varepsilon = J_c^{inter}/J_c^{intra}$ continuously increases with K doping, similar to the case of doped copper-oxides superconductors but found to be different in intrinsic mechanisms. It is found that the FeAs wetting phase widely existing in the grain boundaries is the major factor inhibiting the inter grain critical current. The doping dependence of grain boundary transport properties can be well explained by the model of proximity effect induced SNS Josephson junction. We suggest that the critical current density of iron-based superconducting conductors can be largely enhanced by diminishing the intercalating FeAs phase or varying the electronic properties of grain boundaries through doping.

**CRediT authorship contribution statement**

Yanwei Ma and Chiheng Dong conceived and designed the research. Zhe Cheng and Chiheng Dong performed the experiments and data analysis. Qinghua Zhang and Lin Gu performed the TEM characterization. Satoshi Awaji provided support for transport critical current measurements. Chiheng Dong, Huan Yang and Hai-Hu Wen performed the theoretical calculation. Chiheng Dong, Hai-Hu Wen and Yanwei Ma wrote the manuscript.

**Acknowledgements**

This work is partially supported by the National Key R&D Program of China (Grant Nos. 2018YFA0704200 and 2017YFE0129500), the Strategic Priority Research Program of Chinese Academy of Sciences (Grant No. XDB25000000), the National Natural Science Foundation of China (Grant Nos. 52172275, U1832213, 51861135311, 51721005), Beijing Municipal Natural Science Foundation (Grant No. 3222061), Natural Science Foundation of Shandong Province (Grant No. ZR2021ME061), the Youth Innovation Promotion Association of CAS (Grant No. 2019145), the Key Research Program of Frontier Sciences, CAS (Grant No. QYZDJ-SSW-JSC026), the International Partnership Program of Chinese Academy of Sciences (Grant No. 182111KYSB20160014).

**Data availability statement**

The raw/processed data required to reproduce these findings cannot be shared at this time due to legal or ethical reasons.

# Figures

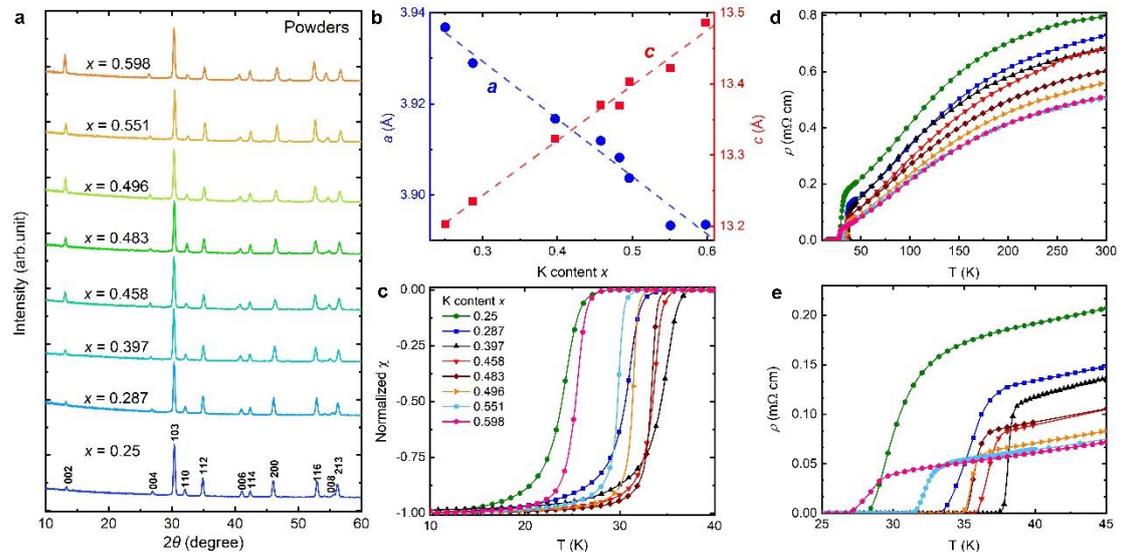

**Fig.1 Crystal structure, susceptibility and resistivity of the $Ba_{1-x}K_xFe_2As_2$ samples.** (**a**), Powders X-ray diffraction patterns. (**b**), Doping dependence of the lattice parameters derived from the Rietveld refinements. Temperature dependence of (**c**) the normalized susceptibility and (**d**) the resistivity. (**e**), enlarged view of the resistivity near the superconducting transition. The legend for (c-e) is shown in (c).

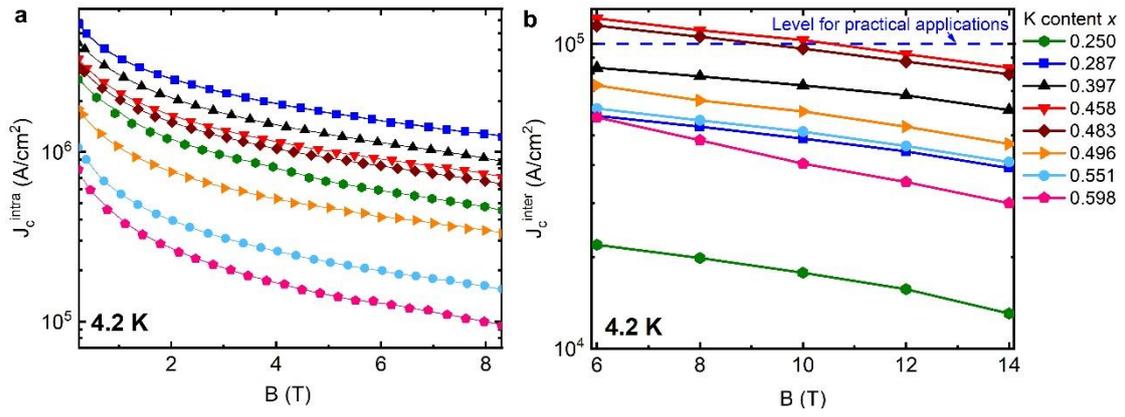

**Fig.2 Critical current density of the Ba$_{1-x}$K$_x$Fe$_2$As$_2$ tapes.** Field dependence of the (a) intra- and (b) inter-grain critical current density. The highest $J_c^{inter}$ of the measured samples is shown here. The blue dashed line in (b) marks the $J_c$ level for practical applications. The legend is shown in (b).

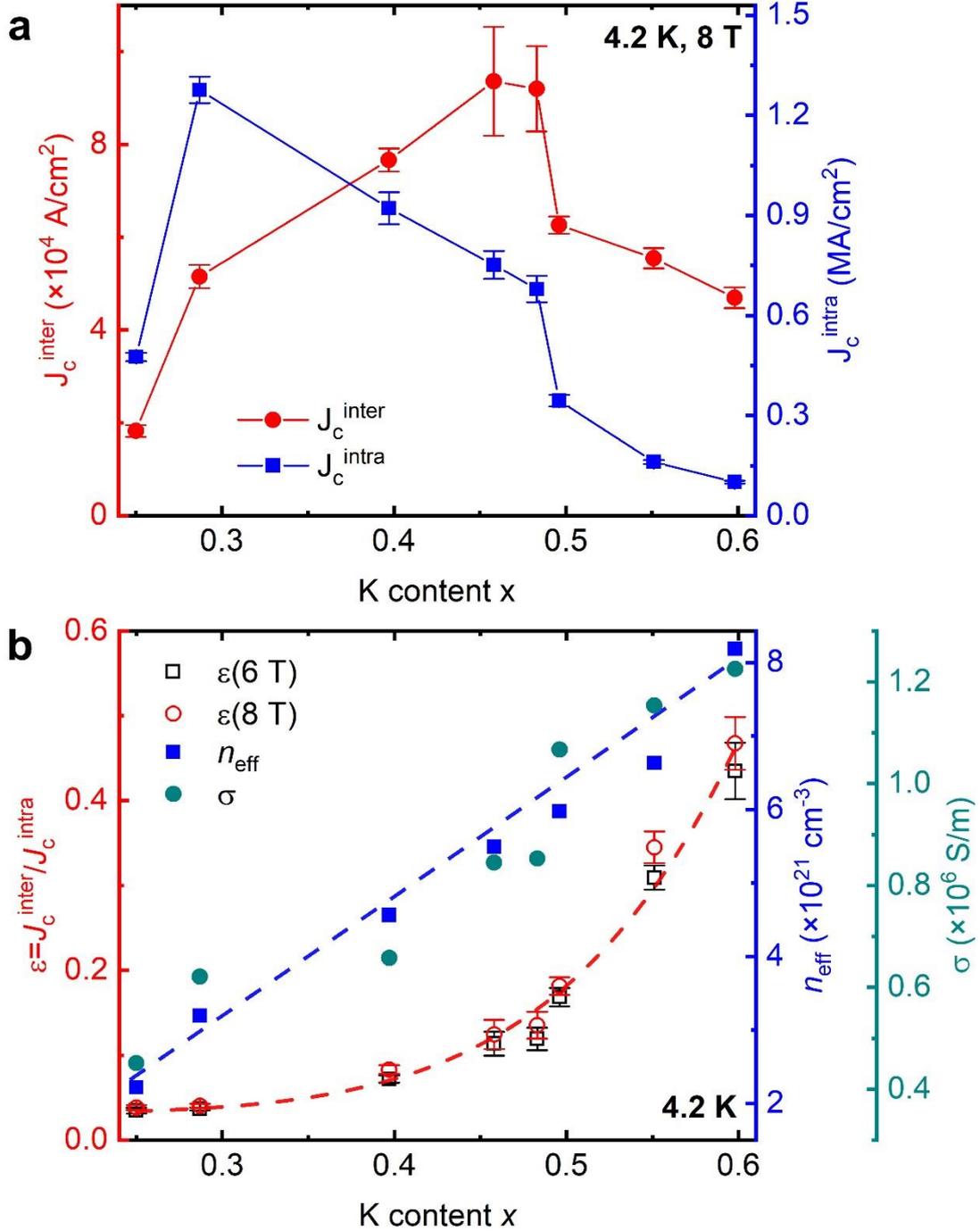

**Fig.3 Doping dependence of the critical current density, charge carrier density and conductivity of the $Ba_{1-x}K_xFe_2As_2$ samples.** (**a**), Inter- and intra-grain critical current densities at 4.2 K as a function of the K content $x$. The average $J_c^{inter}$ of the measured samples is shown here. The error bars denote the standard deviation (*s.d.*) of the measurements. (**b**), Doping dependence of the grain boundary transparency parameter $\varepsilon(4.2\ K)=J_c^{inter}/J_c^{intra}$ at 6 T and 8 T. The error bars are the *s.d.* The effective charge carrier density $n_{eff}$ and the electrical conductivity σ at 50 K as a function of the doping level are also included. The red and blue dashed lines in (b) are guides to the eye.

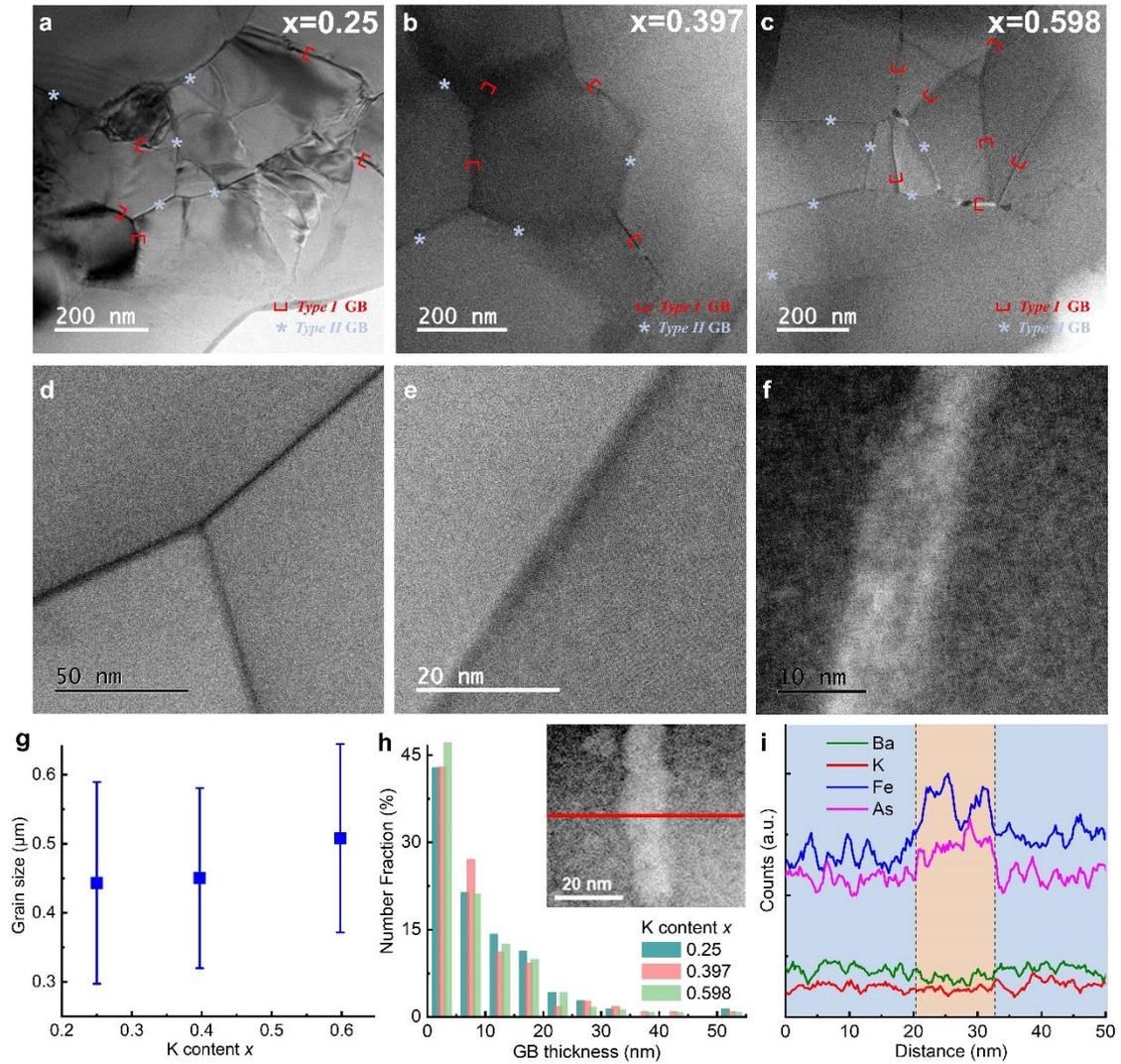

**Fig.4 Microstructural analysis of the GBs.** TEM images for (a) the under-doped ($x$=0.25), (b) optimally-doped ($x$=0.397) and (c) over-doped ($x$=0.598) samples. The ⊓ and * mark the Type I ($t$>5 nm) and type II ($t$<5 nm) GBs, respectively. Enlarged views of the (d-e) type II GBs and (f) type I GBs. (**g**), Grain size as a function of the K content. (**h**), Number fraction of the grain boundary thickness $t$. **i**, EDS line scan across the GB marked by the red line in the inset of (h). The grain-GBs interface is marked by the vertical dashed line, the GBs and the grains are colored by light red and light blue, respectively.

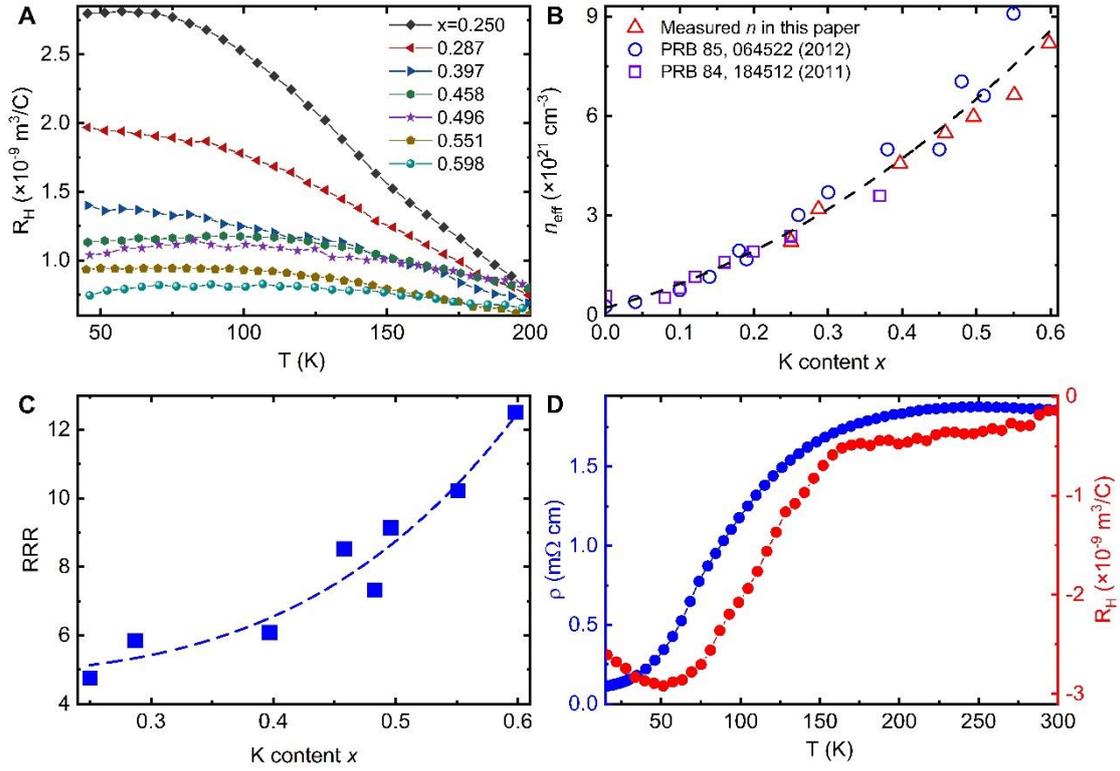

**Fig.5 Electrical measurement results of the $Ba_{1-x}K_xFe_2As_2$ and the FeAs samples.** (**a**), Temperature dependence of the Hall coefficients. (**b**), Doping dependence of the effective carrier density $n_{eff}$ at 50 K. (**c**), Doping dependence of the residual resistance ratio (RRR) defined by the ratio between $\rho(300\ K)$ and the $\rho$ just above $T_c$. The dashed lines in (b, c) are the guides to the eye. (**d**), Temperature dependence of the resistivity (blue) and the Hall coefficient (red) of the grain boundary wetting phase FeAs.

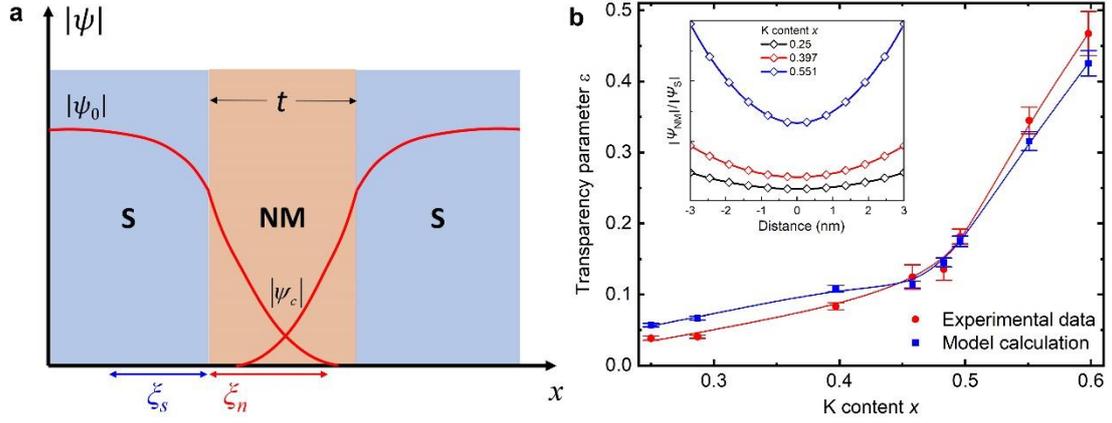

**Fig.6 Theoretical simulation of the proximity-induced SNS junction.** (**a**), A schematic diagram of the SNS junction. The superconducting order parameter $|\Psi_0|$ extends into the GB wetting phase, FeAs (NM). The $|\Psi_0|$ penetrates further into the GB and overlaps with each other with a value of $|\Psi c|$ at the center of the normal phase. (**b**), The GB transparency parameter calculated according to the SNS-based Josephson junction model. The experimental data is also included for comparison. The error bars of the calculated $\varepsilon$ come from the statistical error of the GB thickness. The inset shows the normalized order parameter inside a 6 nm thick NM layer.